# Evaluating the Predictability of Selected Weather Extremes with Aurora, an AI Weather Forecast Model

Qin Huang[1], Moyan Liu[1], Yeongbin Kwon[2], Upmanu Lall[1,3]*

[1]School of Complex Adaptive Systems & Water Institute, Arizona State University, Tempe, AZ, USA.

[2]Department of Civil and Environmental Engineering, Seoul National University, Seoul, Korea.

[3]Department of Earth and Environmental Engineering & Columbia Water Center, Columbia University, New York, NY, USA.

Corresponding author: Qin Huang (qhuang62@asu.edu)

**Key Points:**

- Aurora shows strong 1–7 day skill for tropical cyclones, temperature extremes, and atmospheric rivers, with errors comparable to operational benchmarks.

- At 14–21 days, large-scale circulation remains skillful but extreme intensity collapses, revealing a consistent subseasonal failure mode.

- The practical predictability horizon for actionable extreme-event guidance is ~7–10 days, reflecting intrinsic atmospheric dynamical limits.

**Abstract**

Artificial intelligence (AI) weather foundation models now achieve forecast skill comparable to numerical weather prediction at far lower computational cost, yet their predictability for high-impact extremes across dynamical regimes remains uncertain. We evaluate Aurora using an event-based framework spanning tropical cyclones, freezes, heatwaves, atmospheric rivers, and extreme precipitation at lead times from 1 to 21 days. Aurora demonstrates strong short-range (1–7 day) skill across event types, including competitive tropical cyclone track accuracy and high spatial agreement for temperature and moisture extremes. However, a consistent subseasonal failure mode emerges: while large-scale circulation patterns remain moderately skillful at 14–21 day leads, threshold-based extreme intensity collapses as fields regress toward climatology. This divergence indicates that Aurora retains synoptic-scale dynamical structure but loses surface-impact amplitude beyond ~7–10 days. The practical predictability horizon for deterministic AI extreme-event forecasting therefore remains constrained by intrinsic atmospheric dynamics.

## 1 Introduction

Over the past few years, there has been a blossoming of deep learning-based foundation models for weather forecasting, emerging as a powerful complement to traditional numerical weather prediction (NWP) systems. Early breakthroughs such as Pangu-Weather (Bi et al., 2023) and GraphCast (Lam et al., 2023) demonstrated that data-driven models trained on historical reanalysis can rival or exceed the skill of state-of-the-art NWP for medium-range global forecasting. Since then, the field has advanced rapidly, with a new generation of large-scale models including GenCast (Price et al., 2025), AIFS (Lang et al., 2025), and Aurora (Bodnar et al., 2025). These models differ in architecture (e.g., graph neural networks (GNNs), transformers, diffusion models), training datasets, forecast horizons, and computational design (Table 1), but signal a trend toward data-centric approaches for operational forecasting.

In parallel, traditional NWP systems remain grounded in the physical laws governing atmospheric dynamics and thermodynamics. Operational models such as the European Centre for Medium-Range Weather Forecasts (ECMWF) Integrated Forecasting System (IFS), the National Centers for Environmental Prediction (NCEP) Global Forecast System (GFS), and regional models including the Weather Research and Forecasting (WRF) Model (NCAR) continue to form the backbone of global and regional weather prediction. Their strengths and limitations have been extensively benchmarked against emerging AI forecast models, which show that while NWPs remain the operational standard, challenges persist in accurately predicting the timing, intensity, and spatial structure of certain high-impact weather phenomena (Brotzge et al., 2023; Bouallègue et al. 2024; Allen et al., 2025).

The performance improvement of AI weather models relative to NWP has been rapid and well-documented. Pangu-Weather was the first AI model to cross the IFS threshold, achieving a 5-day Z500 RMSE of 296.7 compared to IFS's 333.7, an approximately 11% improvement, while running more than 10,000 times faster than the operational IFS on a single GPU (Bi et al., 2023). GraphCast extended the benchmark, outperforming IFS-HRES on 90% of 1,380 verification targets and producing a 10-day global forecast in under one minute on a single TPU (Lam et al., 2023). Both two are deterministic models. GenCast, using a diffusion-based ensemble approach, introduced probabilistic skill at scale and became the first AI ensemble system to outperform IFS's operational ensemble (ENS) on 97.2% of 1,320 probabilistic targets, completing a 15-day 50-member ensemble in approximately 8 minutes on a single TPU v5 (Price et al., 2025). Despite these advances, early AI models demonstrated only marginal improvements in tropical cyclone (TC) track prediction relative to operational centers, with comparisons limited to short lead times or reanalysis benchmarks rather than against the operational forecast agencies themselves (Bodnar et al., 2025).

Aurora (Bodnar et al., 2025) represents the current frontier of AI weather forecasting, achieving performance gains across an unprecedented range of forecasting domains. At 0.25° resolution, Aurora outperforms both IFS-HRES and GraphCast on more than 91% of verification targets; at fine-tuned 0.1° resolution, matching the native resolution of the IFS-HRES, it surpasses operational IFS on 92% of targets. Aurora is the first AI model to surpass all official operational TC forecast systems at 1-5 day leads: it outperforms the National Hurricane Center (NHC) by 6% at 1-day lead and 20-25% at 2-5 day leads in the North Atlantic and East Pacific, with average improvements of 18% and 24% in the Northwest Pacific and Australian regions, respectively, outperforming seven operational forecast agencies in total. Aurora's strong performance is attributed to its large parameter count (1.3 billion), transformer-based architecture, and an unusually diverse pretraining dataset spanning ERA5, IFS high-resolution forecasts, HRES-WAM wave analyses, and climate model output, totaling more than one million hours of Earth system data (Bodnar et al., 2025; Zhao et al., 2025).

In many existing studies, direct comparisons among leading AI models have shown that Aurora consistently ranks among the top performers across multiple variables and lead times, particularly for high-impact circulation patterns. Evaluating four AI models across 50 tropical cyclones in five ocean basins, Sahu et al., 2025 found that all AI models achieved mean track errors below 200 km at 96-hour lead, consistently outperforming GFS, IFS, and the UK Met Office Unified Model, with Aurora demonstrating the strongest representation of TC dynamics and thermodynamics among all ML models, an advantage attributed to its multi-source training data. However, while these models often achieve strong track forecasting skill, they tend to underestimate TC intensity compared with observations and NWP systems, potentially a limitation linked to underrepresentation of TC intensity in ERA5 reanalysis training data (Sahu et al., 2025; DeMaria et al., 2025; Gupta et al., 2025) evaluated seven AI weather prediction models against ground-based observations, rain gauges, and satellite data during the South Asian Monsoon. They found that while AI models can capture large-scale monsoon dynamics reasonably well, their forecast errors measured against real observational data were up to ~15-45% larger than errors measured against reanalysis products, indicating that standard reanalysis-centric benchmarks can overstate forecast skill. The study underscores the importance of observation-based validation and event-focused evaluation for regional weather prediction, particularly in data-sparse, high-impact monsoon environments.

**Table 1.** Comparison of Leading AI Weather Forecast Models

|  | Pangu | GraphCast | GenCast | AIFS | Aurora |
|---|---|---|---|---|---|
| Year | 2023 (*Nature*) | 2023 (*Science*) | 2024 (*Nature*) | 2024 (*Arxiv*) | 2025 (*Nature*) |
| Developer | Huawei | Google DeepMind | Google DeepMind | ECMWF | Microsoft |
| Architecture | 3D Earth-Specific Transformer | GNN | Diffusion +GNN | GNN+Transformer | 3D Swin Transformer |
| Parameters | 64M | 36.7M | ~40-60M | ~100M | ~1.3B |
| Training Data | ERA5 (1979-2017) | ERA5 (1979-2017) | ERA5 (1979-2018) | ERA5 (1979-2021)+ ECMWF operational analyses | ERA5 (1979-2020)+ HRES-WAM + IFS HRES T0 + Climate Simulations |
| Resolution | 0.25° | 0.25° | 0.25° | 0.25° | 0.25° (pretrained) / 0.1° (fine-tuned) |

| Max Lead Time | 7 days | 10 days | 15 days | 10 days | 10 days |
| --- | --- | --- | --- | --- | --- |
| Forecast Type | Deterministic | Deterministic | Probabilistic | Deterministic+Probabilistic | Deterministic |
| Speed (Inference) | Seconds per forecast (single GPU) | Less than 1 minute for 10-day forecast (single TPU v4) | ~8 min for 15-day ensemble (single TPU v5) | Minutes (single GPU) | ~0.6s per 6h step (A100 GPU) |

Collectively, these studies point to Aurora as the leading deterministic AI weather foundation model across the broadest range of forecasting domains. Yet existing evaluations focus either on aggregate global performance metrics or on individual event categories in isolation. No study has systematically examined Aurora's predictability across multiple classes of high-impact extremes within a unified event-based framework. This gap is consequential: different extreme weather types are governed by distinct physical mechanisms, operate at different spatial and temporal scales, and present fundamentally different predictability challenges.

This study evaluates Aurora's performance across multiple classes of high-impact weather extremes, including tropical cyclones, freezes, heatwaves, atmospheric rivers, and extreme precipitation. Rather than relying on aggregate global metrics, we analyze a curated set of societally consequential events spanning distinct dynamical regimes and climate contexts. The goal is to diagnose how forecast skill evolves across lead times and physical processes within a state-of-the-art AI foundation model.

## 2 Methods and Data

### 2.1 Event Selection

We adopt a curated event-based experimental design focused on high-impact extreme weather cases spanning distinct geographic regions and dynamical mechanisms. The framework prioritizes physical interpretability and examination of forecast evolution across lead times rather than maximizing sample size.

Events were selected according to four criteria: documented societal impact, meteorological intensity, ERA5 data availability, and representation of distinct dynamical regimes. The inventory includes both in-sample (pre-2020) and out-of-sample (post-2020) cases relative to Aurora's ERA5 training period (1979-2020), enabling assessment of performance within and beyond the training distribution (Table 2).

Prior to model evaluation, ERA5 fields were examined to confirm accurate representation of event timing and magnitude. All event onset, peak, and recovery times are defined using ERA5 following this verification step.

A two-stage framework (Figure 1) is applied consistently across event categories. Stage 1 evaluates lead-time dependence by varying initialization dates while targeting a fixed critical phase (e.g., landfall for tropical cyclones; onset or peak for temperature extremes and atmospheric rivers). Stage 2 examines either sub-daily initialization sensitivity (tropical cyclones) or extended spatial and physical diagnostics (freeze, heatwave, atmospheric river, and heavy precipitation events). This structure enables cross-event comparison while preserving event-specific physical interpretation.

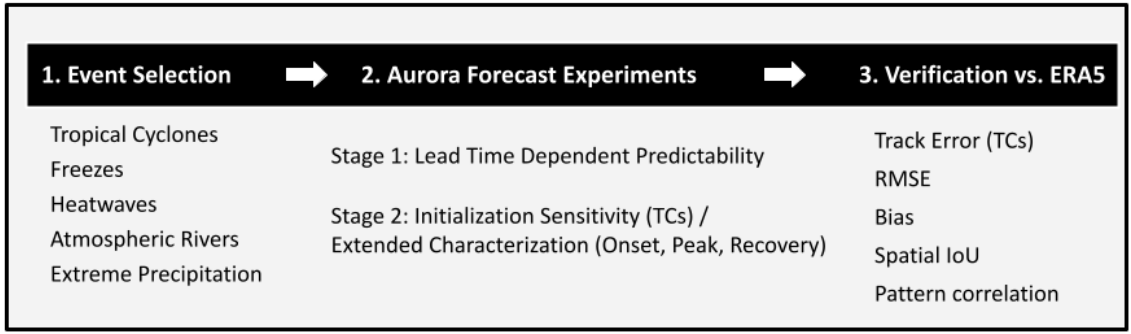

**Figure 1.** Aurora Predictability Study Flowchart.

**Table 2.** Event Inventory.

| Event | Region | Dominant Mechanism | Lead Time Tested |
|---|---|---|---|
| Tropical Cyclones | | | |
| Sandy 2012* | Atlantic, U.S. Northeast (NJ, NY, CT, RI, MA) | TC–extratropical transition with baroclinic amplification | 1, 3, 5, 7 days |
| Amphan 2020* | North Indian Ocean (West Bengal, Odisha, Bangladesh) | Rapid intensification in low-shear, high-OHC environment | |
| Ian 2022 | Atlantic, Gulf of Mexico (Florida, South Carolina) | Rapid intensification over high ocean heat content (Loop Current) in Gulf of Mexico | |
| Hinnamnor 2022 | Western North Pacific (Okinawa, South Korea) | Subtropical ridge steering and midlatitude westerly interaction | |
| Freeze | | | |
| Beast from East 2018* | Western Europe (UK, Ireland, France, Germany, Netherlands) | Ural blocking and SSW-induced polar vortex displacement | 1, 7, 14, 21 days |
| Texas 2021 | Southern United States (Texas) | Stratospheric polar vortex weakening and amplified tropospheric wave pattern | |

| Heatwave | | | |
|---|---|---|---|
| British Columbia 2021 | Western Canada (British Columbia) | Persistent blocking ridge with subsidence amplification | 1, 7, 14, 21 days |
| Southwest European 2023 | Southern and Western Europe (Spain, Italy, Greece, France) | Persistent subtropical ridge and Rossby wave amplification | |
| Atmospheric River | | | |
| Iran 2019* | Southwestern Iran (Khuzestan, Lorestan) | Arabian Sea atmospheric river interacting with midlatitude trough and orography | 1, 3, 5, 7 days |
| California 2022-23 | U.S. West Coast (California) | Successive high-IVT atmospheric rivers associated with extended Pacific jet | |
| Extreme Precipitation | | | |
| Pakistan 2010* | South Asia (Pakistan: Sindh, Punjab) | Enhanced monsoon circulation interacting with midlatitude Rossby wave train (La Niña year) | 1, 3, 5, 7 days |
| Sudan 2020* | East Africa (Sudan: Khartoum, Blue Nile) | Intensified monsoon circulation with Indian Ocean SST anomalies | |
| Western Europe 2021 | Western Europe (Germany, Belgium) | Quasi-stationary cut-off low with extreme moisture convergence | |
| Appalachian 2022 | Eastern Kentucky (Central Appalachia) | Training mesoscale convective system in high-PWAT environment | |

Asterisks (*) denote in-sample events falling within Aurora's ERA5 training period (1979-2020).

## 2.2 Aurora Model Configuration

We use the Aurora 0.25° pretrained model without fine-tuning. The model operates on a 721 × 1440 global grid and produces forecasts at 6-hour intervals (00, 06, 12, and 18 UTC) using a rollout-based prediction scheme.

Forecasts are initialized from ERA5 atmospheric states. Events occurring prior to 2020 fall within the training period, whereas post-2020 events represent fully out-of-sample cases (Table 2). Aurora is treated as a fixed forecasting system; no event-specific tuning is performed. All analyses are diagnostic, focusing on forecast evolution rather than model optimization.

Model inputs include surface variables, 2-m air temperature (T2m), 10-m winds (U10, V10), mean sea level pressure (MSLP), atmospheric fields at 13 pressure levels (temperature, horizontal winds, specific humidity, and geopotential height), and static fields (land-sea mask, soil type, topography). Prognostic outputs include global pressure, wind, temperature, and moisture fields, from which event-specific diagnostics are derived.

## 2.3 Experiment Design

### 2.3.1 Tropical Cyclones (TCs)

Four TC cases spanning three ocean basins were selected to sample a range of synoptic environments and steering-flow regimes: Hurricane Sandy (2012, in-sample), Cyclone Amphan (2020, in-sample) in the North Indian Ocean, Hurricane Ian (2022) in the North Atlantic, and Typhoon Hinnamnor (2022) in the western North Pacific. TC development and driving mechanisms (Blake et al., 2013; IMD, 2020; Bucci et al., 2023; Wang et al., 2023) are briefly noted in Table 2.

The TC framework focuses on track evolution, landfall location, and near-surface intensity, the quantities of primary operational importance. Stage 1 initializes forecasts at 1-, 3-, 5-, and 7-day lead times prior to landfall. For systems whose total lifetime does not permit a full 7-day pre-landfall initialization, the maximum available lead is constrained by storm genesis. Stage 2 examines sub-daily initialization sensitivity by comparing forecasts launched at 00, 06, 12, and 18 UTC on the same lead day, assessing whether initialization-time offsets materially affect track and intensity predictions for these rapidly evolving systems.

Storm centers are identified from the MSLP minimum in Aurora output and compared against IBTrACS best-track positions (Gahtan, J. et al., 2024). Verification metrics include storm-center position error, landfall position error, minimum MSLP bias, and maximum 10-m wind speed bias. Primary diagnostic fields are MSLP and 10-m wind components, from which maximum sustained wind speed is derived.

### 2.3.2 Freezes

Cold-wave predictability is examined through two case studies representing distinct governing dynamics: the February-March 2018 European cold wave ("Beast from the East," in-sample), driven by Ural blocking and sustained Siberian cold-air advection (Huang et al., 2022), and the February 2021 Texas freeze, associated with a weakened stratospheric polar vortex and anomalous Arctic air intrusion over North America (Doss-Gollin et al., 2021).

Stage 1 evaluates freeze detection and onset prediction at 1-, 7-, 14-, and 21-day lead times, spanning the synoptic to subseasonal range. Onset is defined as the first day on which domain-mean T2m drops below 0 °C. Stage 2 characterizes the spatial extent, intensity, and lifecycle evolution of freezing conditions, including onset, peak cold, and recovery, alongside their governing large-scale circulation anomalies. No sub-daily initialization sensitivity tests are performed for freeze events; computational resources are directed instead toward the extended lead times appropriate for these synoptically paced phenomena.

Primary diagnostics employ T2m as the freeze indicator, with threshold levels at 0 °C, -5 °C, and -10 °C. Supporting fields include 850 hPa temperature (T850) for cold-air-mass tracking, 500 hPa geopotential height (Z500) for blocking identification, and MSLP for surface pressure characterization.

### 2.3.3 Heatwaves

Heatwave predictability is assessed using two high-impact events with distinct but analogous large-scale forcing: the June–July 2021 British Columbia heatwave and the August 2023 Southwest European heatwave. The 2021 British Columbia event produced record-breaking temperatures across western Canada under an exceptionally strong and persistent mid-tropospheric blocking ridge that promoted subsidence and anomalous surface warming ([White et al., 2023](#)). The August 2023 Southwest European heatwave brought prolonged extreme temperatures to Spain, France, and surrounding regions under a sustained subtropical high-pressure system and amplified mid-latitude wave pattern ([Magnusson & Napoli, 2023](#)).

Stage 1 evaluates Aurora's ability to predict heatwave occurrence and onset at 1-, 7-, 14-, and 21-day lead times, with onset defined as the first day on which regional T2m exceeds 30 °C. Stage 2 characterizes spatial structure, intensity evolution, and governing circulation, including mid-tropospheric blocking configuration, jet stream position, and warm-air-mass development. Diagnostic fields mirror those used for freeze events with inverted thresholds: T2m (at 30 °C, 35 °C, and 40 °C), T850, Z500, and MSLP. This paired warm/cold event design enables direct comparison of Aurora's predictive behavior across opposite temperature extremes under structurally analogous large-scale dynamical forcing.

### 2.3.4 Atmospheric Rivers (ARs)

Atmospheric rivers (ARs), narrow, transient corridors of anomalously strong poleward moisture transport, are characterized here through spatial structure and landfall intensity rather than point-based tracking metrics. Two high-impact events are analyzed: the 18 March 2019 Iran flood event (in-sample), which affected more than 25 provinces ([Dezfuli, 2020](#)), and the 27 December 2022 U.S. West Coast AR (out-of-sample), which produced widespread flooding and infrastructure damage ([Schubert et al., 2024](#)).

Stage 1 initializes forecasts at 1-, 3-, 5-, and 7-day lead times prior to peak integrated vapor transport (IVT) at landfall (as diagnosed by ERA5) to evaluate Aurora's ability to predict AR timing, landfall position, and intensity from both synoptic and local perspectives, under both in-sample and out-of-sample conditions. Stage 2 examines the spatial coherence of the IVT plume and the synoptic-scale circulation patterns sustaining moisture transport to the affected regions. No sub-daily initialization sensitivity tests are conducted for AR events.

### 2.3.5 Extreme Precipitation Events

Extreme precipitation predictability is evaluated using four high-impact flood events spanning distinct hydroclimatic regimes and both in-sample and out-of-sample periods: the July–August 2010 Pakistan floods (in-sample), which inundated approximately one-fifth of the country under sustained monsoon circulation (Houze et al., 2011); the August 2020 Sudan floods (in-sample), driven by anomalous convective activity during the West African monsoon (FAO, 2020); the July 2021 Western Europe floods (out-of-sample), concentrated in the Ahr and Meuse river basins under a quasi-stationary low-pressure system (C3S, 2021); and the July 2022 Appalachian floods, associated with a slow-moving mesoscale convective system that produced extreme localized accumulations over complex terrain (NWS, 2022).

Precipitation is not a native output variable of the Aurora pretrained model. We therefore use an auxiliary precipitation decoder which maps Aurora's latent atmospheric state representation to precipitation without modifying the pretrained Aurora dynamical core (Lehmann et al., 2025). No additional training or fine-tuning is performed.

Precipitation verification is conducted using the Multi-Source Weighted-Ensemble Precipitation v3 (MSWEP v3) dataset (Wang et al., 2026), which also served as the observational reference during decoder training. Accordingly, verification reflects consistency with the training target rather than fully independent observational validation.

Stage 1 initializes forecasts at 1-, 3-, 5-, and 7-day lead times prior to the observed precipitation peak, targeting the 6-hour accumulation period with the highest regional MSWEP total. This range captures the forecast window most relevant for flood early warning. Stage 2 examines initialization-time sensitivity at 00, 06, 12, and 18 UTC for the 3-day lead, assessing whether sub-daily differences in initialization materially affect precipitation placement and intensity.

**2.4 Forecast Verification and Computational Implementation**

Forecast accuracy is quantified using root-mean-square error (RMSE) and spatial metrics computed against ERA5 reanalysis. We use ERA5 as the verification reference for several reasons. First, ERA5 assimilates many of the same observational data streams (radiosondes, satellites, surface stations) used by operational forecast centers, making it a consistent and comprehensive global baseline (ECMWF, 2019). Second, all forecasts are initialized from ERA5 states, eliminating initialization error as a confounding factor and isolating Aurora's predictive skill degradation with lead time. Third, Aurora was pretrained on ERA5 (1979-2020), making ERA5 the consistent reference for assessing both in-sample generalization and out-of-sample temporal extrapolation.

We also calculate complementary metrics to capture different aspects of forecast quality. The following metrics are computed at event peak times over the regional domain:

*Field accuracy metrics*
Root-mean-square error (RMSE) quantifies the overall temperature field accuracy:

$$RMSE = \sqrt{\frac{1}{N}\sum_{i=1}^{N}(T_{forecast,i} - T_{ERA5,i})^2}$$

where the mean is taken over all grid points in the regional domain.

Mean bias is computed as the spatial average difference:
$$Bias = \overline{T_{forecast}} - \overline{T_{ERA5}}$$
where positive values indicate warm bias and negative values indicate cold bias.

*Pattern metrics*

Spatial pattern correlation measures the structural agreement between forecast and observed fields using the Pearson correlation coefficient computed over all grid points.

$$r = \frac{\sum_{i=1}^{N} (T_{forecast,i} - \overline{T_{forecast}})(T_{ERA5,i} - \overline{T_{ERA5}})}{\sqrt{\sum_{i=1}^{N} (T_{forecast,i} - \overline{T_{forecast}})^2} \sqrt{\sum_{i=1}^{N} (T_{ERA5,i} - \overline{T_{ERA5}})^2}}$$

Pattern correlation values near 1 indicate accurate reproduction of spatial structure, even if temperatures are systematically biased. This metric is particularly useful for assessing whether Aurora captures the location of extreme centers (cold pools, heat domes) independent of their intensity.

*Threshold-based metrics*

For operational early warning, we assess Aurora's skill at predicting the spatial extent of threshold exceedance. Spatial extent is defined as the fraction of grid points exceeding the hazard threshold (T < 0°C for freezes, T > 30°C for heatwaves).

$$Extent = \frac{1}{N} \sum_{i=1}^{N} \mathbb{1}(T_i \leq T_{threshold})$$

To measure spatial overlap between forecast and observed extreme regions, we compute the Intersection over Union (IoU):

$$IoU = \frac{|A_{forecast} \cap A_{ERA5}|}{|A_{forecast} \cup A_{ERA5}|}$$

where $A_{forecast}$ is the set of grid points in the forecast exceeding the threshold, $A_{ERA5}$ is the corresponding set in ERA5, ∩ denotes intersection, and ∪ denotes union. $IoU$ ranges from 0 (no overlap) to 1 (perfect overlap) and provides a stringent test of spatial accuracy: high $IoU$ requires both correct location and correct areal extent of the extreme. This metric is standard in computer vision applications and has recently been adopted in precipitation verification.

For tropical cyclones, track errors are computed as great-circle distances between Aurora-predicted and ERA5 storm centers at each 6-hour timestep, with centers identified as local MSLP minima using an automated tracking algorithm. Landfall position errors are computed as the distance between forecast and observed landfall locations. Intensity errors are quantified as biases in MSLP (hPa) and 10-m wind speed (m/s) at the storm center. For atmospheric rivers, integrated vapor transport (IVT) magnitude errors and spatial displacement metrics are computed. Timing errors for all event types are reported as differences between forecast and ERA5 onset, peak, and recovery times.

It is important to recognize that these metrics capture different aspects of forecast quality and can exhibit different lead-time dependencies. In particular, pattern correlation measures large-scale spatial structure and can remain high even when threshold exceedance skill (IoU) collapses, a behavior we observe systematically at subseasonal leads (§3.2, §3.3). This divergence indicates that Aurora maintains skill at predicting synoptic-scale circulation patterns favorable for extremes (blocking highs, jet stream displacement) while losing skill at predicting surface temperature amplitude. From an operational perspective, high pattern correlation without threshold skill provides limited actionable information for impact-based warnings, as emergency managers require probabilistic estimates of threshold exceedance rather than qualitative indications of anomalous conditions.

**Baseline event detection**. Prior to verification, we performed baseline detection analysis for each event: ERA5 fields were visually inspected and quantitatively assessed to confirm that observed event characteristics are adequately represented in the reanalysis. For tropical cyclones, IBTrACS best-track positions were overlaid on ERA5 MSLP fields to verify cyclone center locations match within 50 km. For temperature extremes, ERA5 T2m fields were compared against station observations from national meteorological services to confirm onset timing and spatial extent. For atmospheric rivers, ERA5-derived IVT was verified against satellite-based estimates. All events passed this baseline check, confirming ERA5 captures the timing, magnitude, and spatial structure of the extreme conditions studied here.

Metrics are evaluated over event-specific spatial domains and temporal windows defined by each event's documented onset, peak, and recovery phases (§2.1), ensuring that verification is anchored to the physically relevant period rather than arbitrary fixed intervals. For each event category, verification is applied to the primary diagnostic fields identified in §2.3: MSLP and 10-m wind speed for tropical cyclones; T2m, T850, and Z500 for freeze and heatwave events; IVT for atmospheric rivers. Where applicable, scalar track-based errors (storm-center displacement, landfall position offset) are reported alongside gridded metrics to preserve physically interpretable measures of forecast quality. While direct comparison against operational NWP forecasts (IFS, GFS) would provide additional context, such comparisons are beyond the scope of this diagnostic study, which focuses on characterizing Aurora's intrinsic predictability limits rather than competitive benchmarking.

**Computational implementation.** All forecasts were generated on a single NVIDIA A100 GPU (80 GB memory); computation time and memory usage were documented for each run. Modern operational NWP systems, by contrast, rely on large-scale supercomputing infrastructure to support high-resolution deterministic and ensemble forecasts ([Bauer et al., 2015](Bauer et al., 2015)). Recent assessments further note that machine-learning–based weather models can be executed with substantially lower computational cost than traditional NWP systems, which remain dependent on high-performance computing clusters for data assimilation and ensemble generation ([Bouallègue et al., 2024](Bouallègue et al., 2024)). Although these systems are not directly comparable—single-member

AI inference versus full operational ensemble forecasting—the contrast highlights the computational efficiency associated with AI-based forecast inference.

## 3 Results and Discussion

### 3.1 TCs

Aurora produced useful deterministic track guidance for several TC cases, but skill varied strongly by event and lead time. Short-range (1-, 3- day) track forecasts were generally the most reliable, while longer leads (5-, 7- day) showed substantial and event-dependent degradation. The four case studies (Hurricane Sandy 2012, Cyclone Amphan 2020, Hurricane Ian 2022, and Typhoon Hinnamnor 2022) illustrate this pattern and expose specific failure modes relevant for operational use. Forecast generation for the four TC cases across all tested lead times and initialization times required ~6 minutes of GPU time on a single NVIDIA A100 (see §2.4 and SI 1).

3.1.1 Lead-Time Dependent Track Skill

Track error increases with lead time, but with large storm-to-storm variability (Table 3; Fig. 2). Aggregating across the four cases, mean track errors are smallest at 1 day (tens of km) and grow to $O(10^2)$ km by 5–7 days. Examples from individual cases highlight the range of behavior:

- **Hurricane Sandy (2012)**. Useful short-range guidance (1-day mean error ≈39 km) but larger degradation by 3–7 days (3-day mean ≈77 km; 7-day ≈134 km), consistent with growing uncertainty in steering and midlatitude interactions.
- **Cyclone Amphan (2020)**. Intermediate behavior, with modest 1–3 day errors (≈23–31 km) but increased spread at longer leads.
- **Hurricane Ian (2022)**. Strong short-range performance: 1-day mean track error ≈13 km and excellent landfall guidance (3-day landfall error reported as 0 km in the Aurora-ERA5 comparison). Ian's performance illustrates Aurora's capability to capture track evolution when steering flow is well defined.
- **Typhoon Hinnamnor (2022)**. A clear failure mode for systems undergoing midlatitude interaction and recurvature: several Stage-2 runs produced very large final track errors (hundreds to >600 km), and one reported landfall/location errors exceeding $10^3$ km. Diagnostic inspection indicates misplacement/timing errors in the midlatitude steering trough and premature phasing with the westerlies as plausible causes.

These results show Aurora is valuable for 1–3 day deterministic track guidance for well-behaved systems, but deterministic skill degrades and becomes highly event dependent at synoptic-to-extended leads.

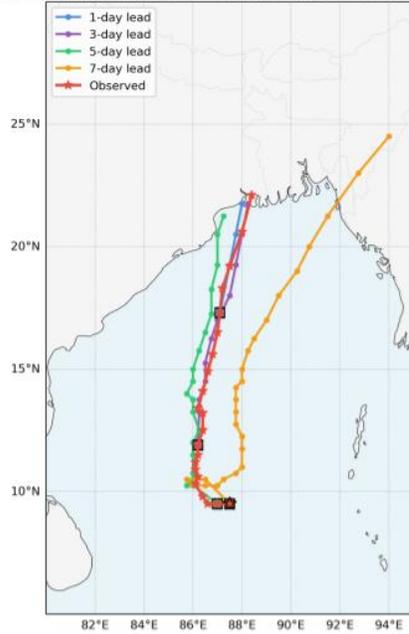
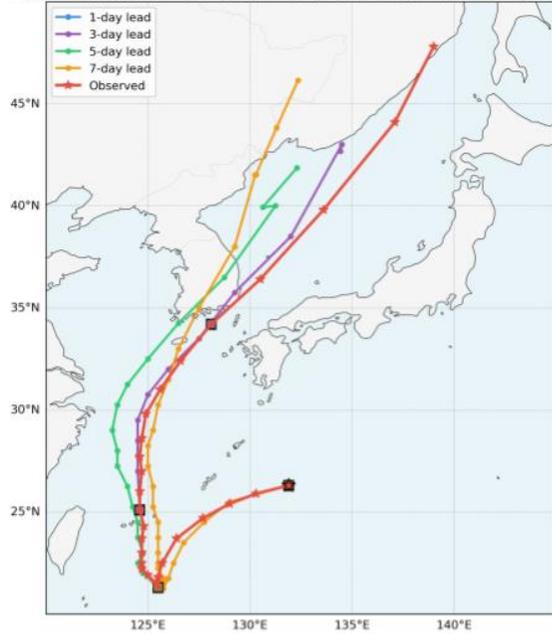
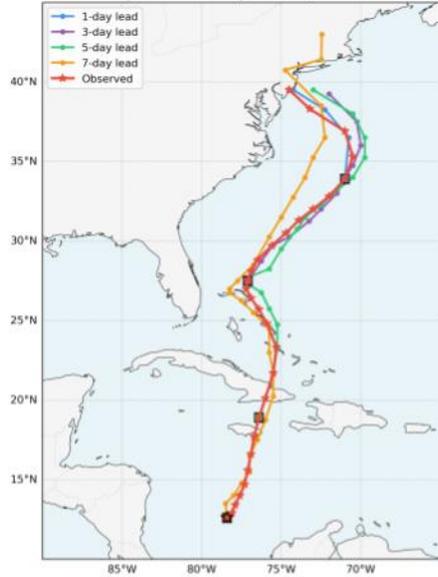
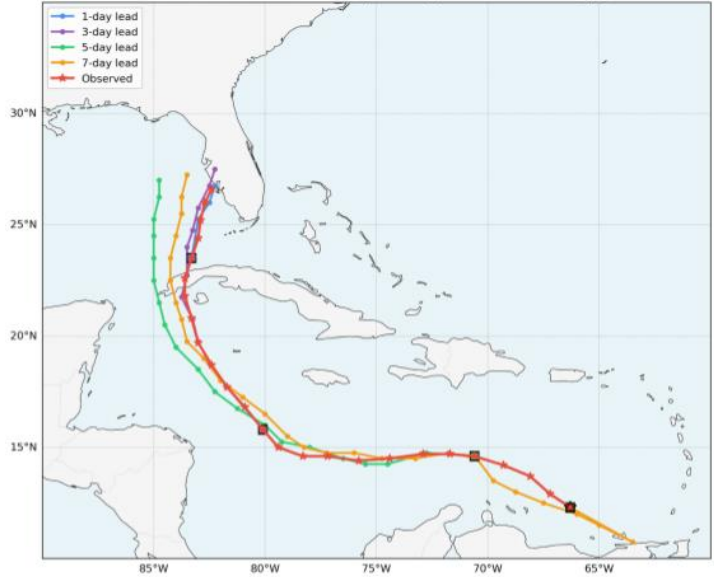

**Figure 2.** Track forecast for 4 TC events for 1-, 3-, 5-,7- lead days.

**Table 3.** TC Stage 1 Lead Time Performance.

| Event | 1-day Mean Error (km) | 3-day Mean Error (km) | 5-day Mean Error (km) | 7-day Mean Error (km) | 3-day Landfall Error (km) |
|---|---|---|---|---|---|
| Sandy 2012* | 38.6 | 77.3 | 63.6 | 134.2 | 243.2 |
| Amphan 2020* | 22.6 | 30.6 | 60.0 | 254.6 | 27.8 |
| Ian 2022 | 13.5 | 35.3 | 98.7 | 74.0 | 0.0 |
| Hinnamnor 2022 | Tracker failed | 132.1 | 155.7 | 170.7 | 1689.7 |

To contextualize these results, official National Hurricane Center (NHC) verification statistics indicate that recent Atlantic basin mean official track errors are on the order of ~30-40 km at 24 h, ~60-80 km at 48 h, and ~90-120 km at 72 h lead time (NHC, 2022). Aurora's short-range track errors for Ian (13.5 km at 1-day; 35.3 km at 3-day) and Amphan (22.6 km at 1-day; 30.6 km at 3-day) fall within the range of recent operational climatological errors, while Sandy's 3-day error (77.3 km) is comparable to typical 72-hour official forecast errors. However, Aurora's performance at longer leads (5–7 days), particularly for recurving systems such as Hinnamnor, degrades beyond typical operational deterministic guidance, highlighting sensitivity to midlatitude interaction. These comparisons are illustrative rather than definitive, as direct paired forecasts against operational NWP systems were not performed in this study.

3.1.2 Initialization-Time Sensitivity

Stage-2 experiments (00/06/12/18 UTC inits at 3-day lead) show modest initialization sensitivity for most cases, but with notable exceptions. The standard deviation of 72-hour track errors across init times is typically $O(10^1)$ km for Sandy, Ian, and Amphan, indicating limited sub-daily dependence. Hinnamnor exhibits large init-sensitivity (std > 100km), reflecting the sharp dependence of recurvature timing on small changes to the vortex-trough phasing. Where init-time differences exist (for example, Ian's 18 UTC initialization outperformed 12 UTC), they likely reflect a combination of sampling of the diurnal/observational cycle in ERA5 and flow-dependent sensitivity of convective organization or steering flow. Larger samples are needed to determine whether systematic init-time preferences exist.

3.1.3 Intensity Biases and Limitations

Across our four cases Aurora exhibits event- and lead-dependent intensity biases; with the current small sample we do not claim a robust basin-dependent pattern. In our runs, some Atlantic cases (Ian, Sandy) show modest positive MSLP biases (weaker storms relative to ERA5) and negative 10-m wind biases at several leads while other cases (Amphan, Hinnamnor) show differing signs or magnitudes of bias.

Because intensity errors are sensitive to small errors in storm structure and to the particular decoder/diagnostic used to derive 10-m winds from Aurora's prognostic fields, these results should be interpreted cautiously. The observed intensity errors (up to ~10 hPa or several m s$^{-1}$ in some runs; see SIx) are large enough to be operationally relevant and suggest that post-processing (e.g., bias correction or model-output statistics) or explicit ensemble approaches would be necessary before Aurora could be used for deterministic intensity guidance.

3.1.4 Operational Implications and Caveats

Aurora demonstrates operationally useful short-range track guidance (1–3 day leads) for several of the examined tropical cyclones when verified against IBTrACS best-track positions. In multiple cases, mean track errors at short lead times fall within ranges typically considered actionable for deterministic guidance. Combined with rapid single-GPU inference, this suggests Aurora could serve as a complementary tool for real-time monitoring or ensemble design.

However, important failure modes remain. Systems undergoing extratropical transition or sharp recurvature, such as Typhoon Hinnamnor, produced very large track errors in deterministic runs, highlighting sensitivity to midlatitude steering-flow interactions and phasing errors. These results indicate that deterministic ML forecasts may require hybrid strategies, such as ensemble generation, coupling with dynamical models, or targeted corrections in recurvature scenarios, for reliable operational deployment.

It is also important to emphasize that these conclusions are based on four case studies and IBTrACS best-track verification. While the experiments reveal both strengths and limitations, they do not substitute for large-sample intercomparison against operational NWP ensembles. Direct comparisons with IFS or GFS were not performed here; therefore, contextualizing Aurora's performance relative to operational systems requires reference to published track-error climatologies or dedicated paired forecast experiments in future work.

**3.2 Freezes**

Freeze predictability exhibits a clear transition between medium-range (1–7 days) and subseasonal (14–21 days) lead times. Results from the Beast from the East (2018, in-sample) and Texas 2021 freeze (out-of-sample) demonstrate Aurora's capability for operational freeze prediction at weekly timescales, but highlight substantial limitations at longer leads. Complete forecast generation for both events across all lead times required approximately 10 minutes of GPU time, with 21-day forecasts consuming the most computational resources (~1 minute each).

3.2.1 Lead-Time Skill Degradation

At 1-day lead, both events show strong spatial skill (Table 5). For Beast from the East, pattern correlation reaches 0.994, spatial extent is 68.5%, and IoU is 0.979, indicating near-complete overlap of the freeze footprint. RMSE is 7.03°C with a minimal warm bias of +0.19°C. For Texas 2021, pattern correlation is similarly high (0.976) and IoU reaches 0.952, demonstrating accurate spatial placement of the cold anomaly. However, RMSE is larger (19.76°C), reflecting amplitude and timing sensitivity during rapid temperature decline. The small mean bias (+0.24°C) indicates little systematic warm offset at short lead. Together, these results show that Aurora reproduces the spatial structure of extreme cold reliably at 1-day lead, even when amplitude errors are non-negligible.

Skill remains robust at 7-day lead. Beast maintains high pattern correlation (0.945) with IoU of 0.902 and modest bias (+0.96°C). Texas shows strong pattern correlation (0.912) and IoU of 0.831, though warm bias increases to +4.96°C. At weekly leads, Aurora captures the occurrence and spatial structure of freeze events, though amplitude underestimation begins to emerge.

At 14–21 day leads, skill deteriorates sharply, but in a diagnostically informative way. For Beast from the East, 14-day pattern correlation remains moderately high (0.875), yet spatial extent collapses to 10.6% and IoU falls to 0.157. RMSE appears low (2.82°C) because the forecast regresses toward climatological values rather than reproducing the extreme amplitude. Warm bias increases to +6.88°C, indicating substantial underestimation of cold intensity. By 21 days, pattern correlation declines to 0.573, spatial extent remains low (19.9%), and IoU remains weak (0.295).

Texas 2021 shows even stronger amplitude collapse at long leads. At 14 days, spatial extent falls to 0.2% (IoU 0.002) despite moderate pattern correlation (0.728) and strong warm bias (+11.62°C). At 21 days, pattern correlation rebounds to 0.852, but spatial extent remains negligible (3.2%, IoU 0.038) and bias remains large (+7.15°C).

This combination of moderate pattern correlation with near-zero threshold skill indicates that Aurora preserves aspects of the large-scale circulation pattern while losing the amplitude of surface temperature extremes. The model retains synoptic-scale structural information but fails to sustain thermodynamic intensity at subseasonal leads.

**Table 5.** Temperature Extremes Performance Summary (Peak Phase, Stage 1)

| Event | Lead Time | T2m RMSE (°C) | T2m Bias (°C) | Pattern Corr | Spatial Extent (%) | IoU (0°C) |
|---|---|---|---|---|---|---|
| Freeze | | | | | | |
| Beast from East 2018* | 1-day | 7.03 | +0.19 | 0.994 | 68.5 | 0.979 |
| | 7-day | 7.40 | +0.96 | 0.945 | 66.1 | 0.902 |
| | 14-day | 2.82 | +6.88 | 0.875 | 10.6 | 0.157 |
| | 21-day | 4.74 | +3.58 | 0.573 | 19.9 | 0.295 |
| Texas 2021 | 1-day | 19.76 | +0.24 | 0.976 | 78.7 | 0.952 |
| | 7-day | 9.29 | +4.96 | 0.912 | 68.7 | 0.831 |

|  | 14-day | 7.81 | +11.62 | 0.728 | 0.2 | 0.002 |
|---|---|---|---|---|---|---|
|  | 21-day | 13.60 | +7.15 | 0.852 | 3.2 | 0.038 |
| Heatwave | | | | | | |
| Southwest Europe 2023 | 1-day | 4.98 | -2.20 | 0.950 | 57.1 | 0.785 |
|  | 7-day | 2.02 | -2.91 | 0.856 | 57.8 | 0.739 |
|  | 14-day | 1.39 | -3.75 | 0.738 | 50.1 | 0.625 |
|  | 21-day | 3.56 | -6.44 | 0.793 | 1.3 | 0.018 |
| British Columbia 2021 | 1-day | 19.06 | -1.00 | 0.942 | 87.1 | 0.938 |
|  | 7-day | 7.59 | -4.96 | 0.914 | 57.0 | 0.625 |
|  | 14-day | 2.85 | -10.63 | 0.800 | 2.1 | 0.024 |
|  | 21-day | 6.04 | -8.26 | 0.856 | 17.6 | 0.193 |

Operational records show that multiple winter weather products and advisories were issued in the days leading up to the peak of the February 2021 Texas cold outbreak, including Winter Weather Advisories on 11 February and expanded Storm Warnings by 13–14 February as Arctic air progressed southward across the region (NWS, 2021). NOAA's official retrospective similarly characterizes the event as a prolonged and historically severe cold outbreak that affected the entire state and set numerous low-temperature records (NCEI, 2021). These operational forecasting activities are broadly consistent with Aurora's strong 1- and 7- day performance for freeze events.

The sharp degradation of freeze extent skill beyond 14 days in Aurora's forecasts is consistent with the intrinsic predictability limits of midlatitude weather systems. Studies of midlatitude atmospheric dynamics suggest that the practical deterministic predictability horizon for large-scale weather phenomena is on the order of ~10 days, beyond which uncertainties in the evolution of Rossby waves and blocking patterns grow rapidly (Sun et al., 2019). Although operational blocking event forecasts improve skill relative to random forecasts at medium range, the inherent instability of blocking and associated downstream flow structures with increasing lead time contributes to diminishing forecast amplitude and threshold skill. Thus, the divergence we observe—moderate pattern correlation persisting at long leads despite collapsed threshold

skill—reflects these fundamental dynamical predictability constraints rather than solely model-specific deficiencies.

3.2.2 Physical Mechanism Representation

At 1–7 day leads, Aurora reproduces the key synoptic structures driving both freeze events. For Beast from the East, forecasts capture the Scandinavian blocking ridge (Z500 anomalies >200 m), southward jet displacement, and sustained cold-air advection into Europe (SI2). For Texas 2021, forecasts reproduce the southward Arctic air intrusion, associated surface high-pressure ridge, and large-scale circulation anomalies consistent with polar vortex displacement (SI3). At 14–21 days, these structures weaken systematically. Blocking amplitudes decrease, cold-air anomalies diminish, and spatial coherence degrades. This behavior is consistent with the known predictability limits of blocking regimes, which are highly sensitive to upstream Rossby wave phase errors that amplify with lead time.

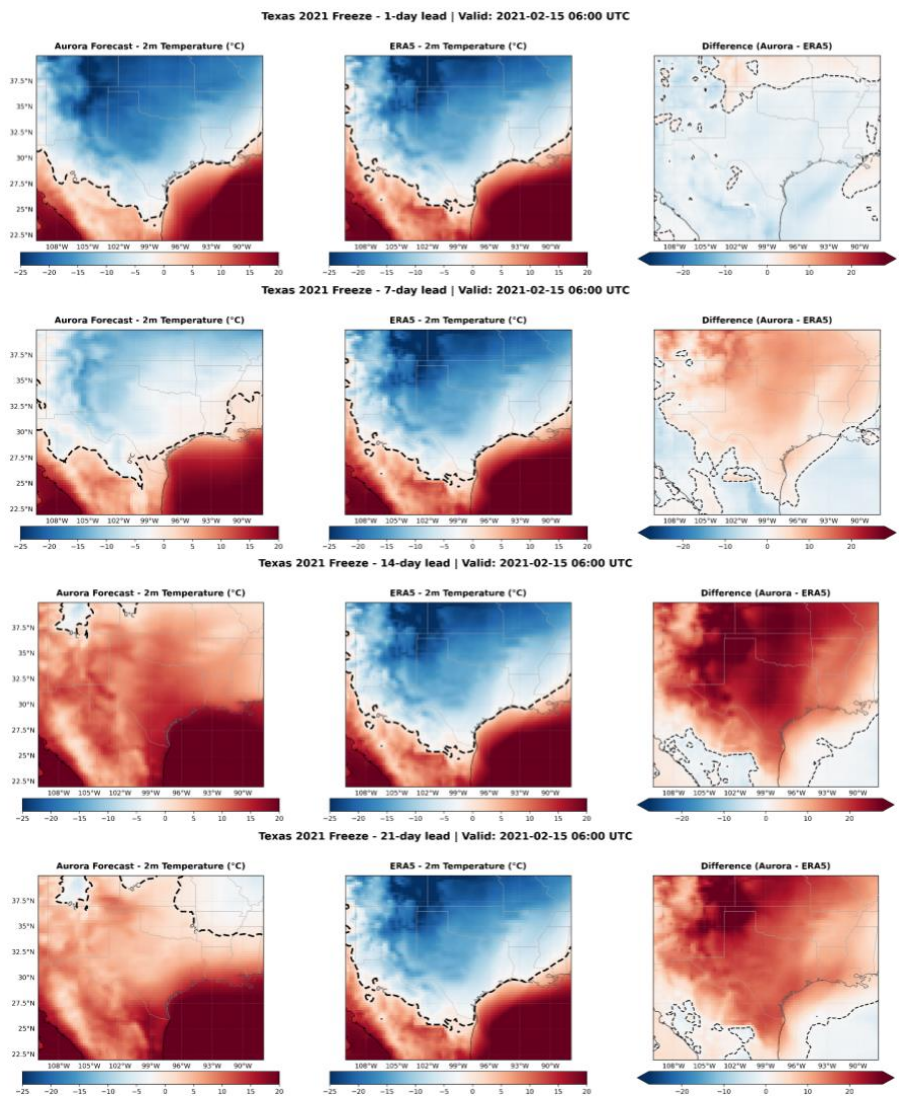

**Figure 3.** Texas Freeze 2021 forecasts (1, 7, 14, 21 lead) at Peak for 2-meter temperature.

3.2.3 Operational Implications

Aurora provides strong operational freeze guidance at 1–7 day leads, with IoU values between 0.83-0.98 and pattern correlations exceeding 0.90. Short-lead biases are small and potentially correctable through post-processing. However, subseasonal performance collapses as IoU drops to near zero by 14–21 days despite retained pattern correlation. This indicates that Aurora can identify circulation regimes favorable for extremes but cannot reliably maintain surface amplitude at these horizons. While this structural skill may still inform ensemble or regime-based outlooks, deterministic subseasonal freeze forecasts remain unreliable without substantial calibration.

**3.3 Heatwaves**

Heatwave predictability was assessed using two out-of-sample events: the August 2023 Southwest European heatwave and the June 2021 British Columbia heatdome. Results reveal markedly different behavior compared to freeze events, with stronger spatial extent skill at short-to-medium leads but similar subseasonal collapse, providing insight into Aurora's treatment of positive versus negative temperature extremes. Similar to freeze events, forecast generation for both heatwave events across all four lead times (1, 7, 14, 21 days) required approximately 10 minutes of GPU time on a single NVIDIA A100.

3.3.1 Short-to-Medium Lead Performance

At 1-day lead, both events show strong spatial performance (Table 5). For the Southwest Europe 2023 heatwave, Aurora forecasts captured 57.1% spatial extent above the 30 °C threshold at peak phase with an IoU of 0.785 and pattern correlation of 0.950. The British Columbia 2021 heat dome showed even stronger 1-day performance (87.1% extent, IoU 0.938, pattern corr 0.942), though with larger RMSE (19.06 °C) likely reflecting the event's sharp temperature gradients.

At 7-day lead, useful guidance persists: Southwest Europe maintains spatial extent (57.8%, IoU 0.739) and a high pattern correlation (0.856), while British Columbia retains moderate extent (57.0%, IoU 0.625) and pattern correlation (0.914). The tendency toward systematic cold bias (e.g., -2.91 °C and -4.96 °C) indicates underprediction of amplitude even when spatial structure is well captured.

At short to medium leads, Aurora's performance aligns with signals present in operational numerical forecasts. For the August 2023 Southwest European heatwave, ECMWF extended-range ensemble means showed anomalously high 2-metre and 850 hPa temperatures several weeks before the event peak, with weekly anomaly forecasts ~3 °C above model climate for parts of southern France starting around 7 August. The ensemble mean crossed the 99th percentile of the model climate distribution approximately six days before the core 22–24 August heatwave period, indicating that global models had a robust heatwave signal well ahead of the event. However, due to wind direction forecast errors along the Atlantic coast, medium-range predictions underestimated extreme surface temperatures in some areas despite correct upper-air signals. In this context, Aurora's ability to capture spatial structure and timing at 1–7 day leads reflects the presence of strong, predictable circulation anomalies that operational extended-range systems also signaled (Magnusson & Napoli, 2023; Magnusson, 2023).

### 3.3.2 Subseasonal Degradation

Predictability degrades at 14–21 day leads, but in a manner that differs from freeze events. At 14-day lead, the European heatwave still exhibits moderate spatial extent (50.1%, IoU 0.625) and decent pattern correlation (0.738), though timing errors grow substantially. For British Columbia, the 14-day forecast shows a near collapse of spatial extent (2.1%, IoU 0.024) despite maintained pattern correlation (0.800) and large cold bias (-10.63 °C), reflecting regression toward climatology. RMSE values at 14 days can appear deceptively low because predicted fields are closer to seasonal climatology than the actual extreme magnitude.

By 21 days both events show near collapse of threshold skill, with the European event at 1.3% extent (IoU 0.018) and British Columbia at 17.6% (IoU 0.193) although pattern correlations remain surprisingly high. These patterns mirror the freeze results and indicate that, at subseasonal lead times, Aurora captures large-scale circulation patterns associated with heat extremes but fails to maintain accurate surface temperature amplitude.

### 3.3.3 Freeze-Heatwave Skill Asymmetry

Comparing heatwaves and freezes reveals systematic differences at subseasonal scales. Heatwave IoU tends to decline more rapidly at mid-range leads (e.g., British Columbia at 14 days) than some freeze cases, possibly reflecting differences in feedback processes (such as soil moisture and land–atmosphere coupling) and the dynamical persistence of the governing circulation regimes. Aurora exhibits systematic cold bias for heatwaves across leads, in contrast to warm bias for freezes at long leads, suggesting asymmetric amplitude underestimation likely tied to training data and loss functions that favour regression toward the mean surfaces.

Although extended-range ensemble forecasts exhibited a signal for the 2023 Southwest European heatwave well ahead of onset, the ability to translate those signals into accurate surface extremes diminished at subseasonal leads, consistent with the collapse of Aurora's threshold skill at 14–21 days (Figure 4). This pattern, early indication of favorable circulation but lost amplitude/extreme detail at longer leads, mirrors the forecast evolution seen in state-of-the-art operational systems and points to intrinsic limitations of deterministic subseasonal predictability.

### 3.3.4 Operational Implications

To assess the predictive skill of Aurora for the December 2022 California atmospheric river (AR) event, we evaluate the integrated vapor transport (IVT) field at the ERA5-diagnosed peak landfall time (27 December 2022, 12:00 UTC) across increasing forecast lead times. Skill is quantified using root-mean-square error (RMSE), domain-mean bias, and spatial pattern correlation relative to ERA5 over both a Pacific outlook domain and the California regional domain (Figure 5 and SI4). Forecast generation for both AR events across all four lead times (1, 3, 5, 7 days) required approximately 2 minutes of GPU time.

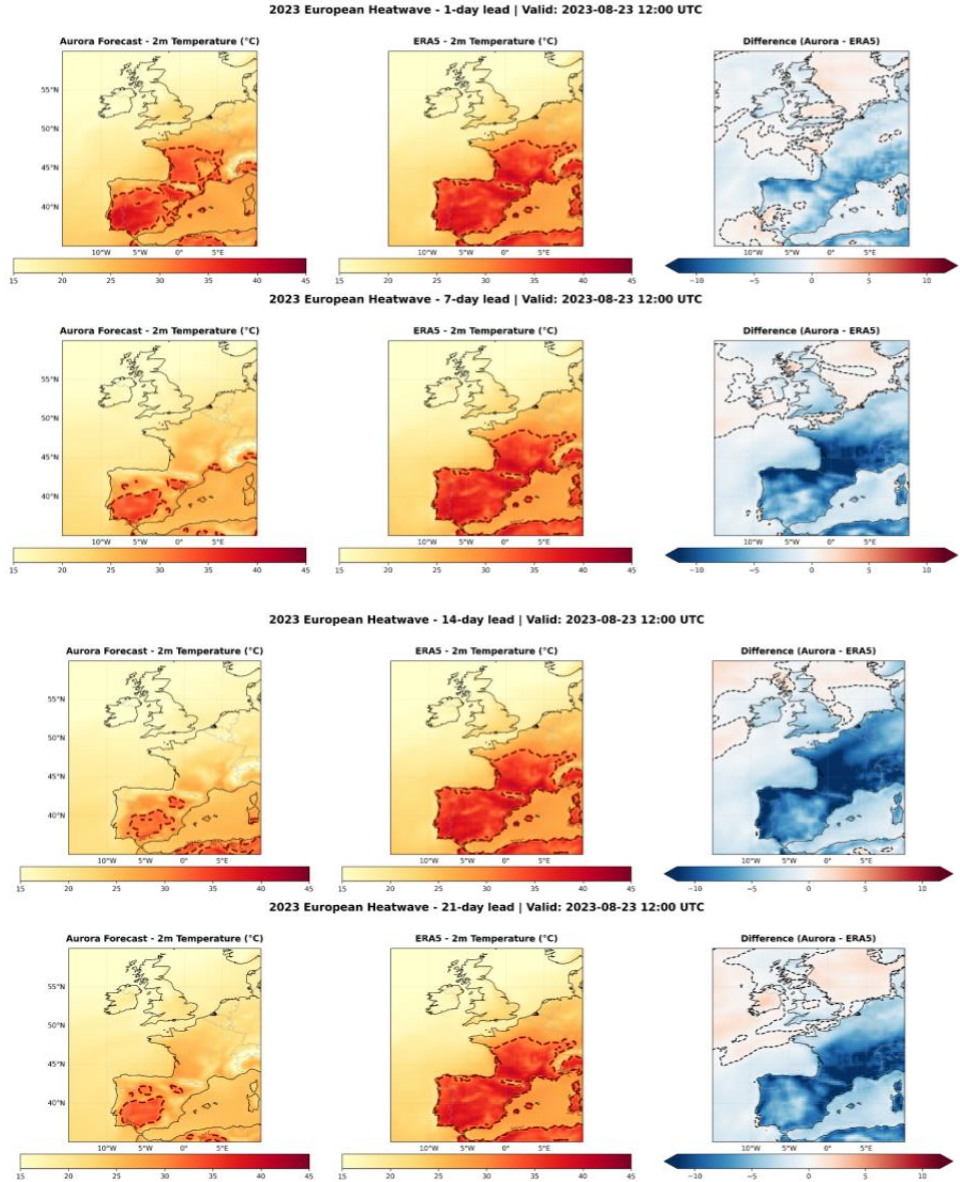

**Figure 4.** European Heatwave 2023 forecasts (1, 7, 14, 21 lead) at Peak for 2-meter temperature.

**3.4 ARs**

To assess the predictive skill of Aurora for the December 2022 California atmospheric river (AR) event, we evaluate the integrated vapor transport (IVT) field at the ERA5 diagnosed peak landfall time (27 December 2022, 12:00 UTC) across increasing forecast lead times. Skill is quantified using root-mean-square error (RMSE), domain-mean bias, and spatial pattern correlation relative to ERA5 over both Pacific outlook domain and CA regional domain (Figure 5 and SI5). Forecast generation for both AR events across all four lead times (1, 3, 5, 7 days) required approximately 2 minutes of GPU time.

3.4.1 Lead-Time Dependent Skill

At short lead times (1-day), Aurora reproduces the AR structure with very high fidelity, low RMSE, and minimal mean bias (Table 6). The forecast accurately captures the filamentary IVT core and its coupling with the mid-latitude Z500 trough, indicating that both moisture transport and large-scale dynamical forcing are well represented at short range.

Skill degrades with increasing lead time. By 3 days, RMSE nearly doubles and spatial correlation declines modestly, although the overall plume orientation and landfall location remain coherent. At 5 days, structural errors become more pronounced and pattern correlation drops to 0.84, reflecting increasing displacement and amplitude errors in the IVT core and associated trough positioning. At 7 days, forecast degradation becomes substantial and spatial correlation decreases to 0.74, indicating partial loss of coherent plume structure. While the large-scale circulation pattern remains broadly recognizable, the IVT maximum weakens and shifts, producing larger mismatches relative to ERA5.

Overall, Aurora demonstrates short-range skill in capturing AR geometry and intensity, with gradual degradation characteristic of nonlinear error growth in mid-latitude moisture transport systems. Spatial structure persists more robustly than amplitude at intermediate leads (3–5 days), suggesting that gross circulation features are easier to predict than exact moisture strength and placement.

| Event | Lead Time | IVT RMSE (kg m$^{-1}$s$^{-1}$) | IVT Bias (kg m$^{-1}$s$^{-1}$) | IVT Pattern Corr |
|---|---|---|---|---|
| Iran AR 2019* | | | | |
| Synoptic Outlook Domain | 1-day | 26.24 | -2.40 | 0.98 |
| | 3-day | 54.20 | -3.21 | 0.94 |
| | 5-day | 87.94 | -6.58 | 0.84 |
| | 7-day | 111.15 | -3.11 | 0.74 |
| Regional Impact Domain | 1-day | 27.97 | 5.34 | 0.98 |
| | 3-day | 46.47 | 0.95 | 0.96 |
| | 5-day | 155.95 | -57.50 | 0.58 |
| | 7-day | 169.08 | -86.11 | 0.59 |
| California AR 2022 | | | | |
| Synoptic Outlook Domain | 1-day | 32.89 | -4.16 | 0.98 |
| | 3-day | 61.93 | -7.03 | 0.94 |
| | 5-day | 106.35 | -8.58 | 0.84 |
| | 7-day | 133.00 | -13.25 | 0.74 |
| CA Regional Domain | 1-day | 75.31 | -29.37 | 0.95 |
| | 3-day | 137.54 | -38.88 | 0.85 |
| | 5-day | 140.37 | -29.79 | 0.85 |
| | 7-day | 379.91 | -314.94 | 0.50 |

**Table 6.** Atmospheric Rivers Performance Summary

3.4.2 Local Level Forecast Skill (California Region)

At 1-day lead time, Aurora maintains strong local skill and high spatial pattern correlation (0.95) within California (Table 6). Although the overall IVT plume structure is well captured, the mean bias is more negative (–29.4 kg m$^{-1}$ s$^{-1}$) than in the broader domain, indicating a systematic underestimation of IVT magnitude at landfall. This suggests amplitude errors emerge earlier at the regional scale than in the synoptic-scale assessment.

Skill declines more rapidly over California than over the larger Pacific domain as lead time increases. By 3–5 days, RMSE increases and pattern correlation decreases, reflecting growing displacement and structural errors within the AR core over the state. At 7 days, forecast degradation is pronounced at the local scale, with substantial breakdown in the representation of IVT structure over California.

In summary, the regional analysis reveals that local landfall skill deteriorates more rapidly than large-scale plume structure. While synoptic coherence persists at intermediate lead times, regional impacts become disproportionally more uncertain once the AR intersects the California coastline. This lead-dependent amplification of local error highlights the sensitivity of landfall forecasts to subtle changes in the moisture transport pathway.

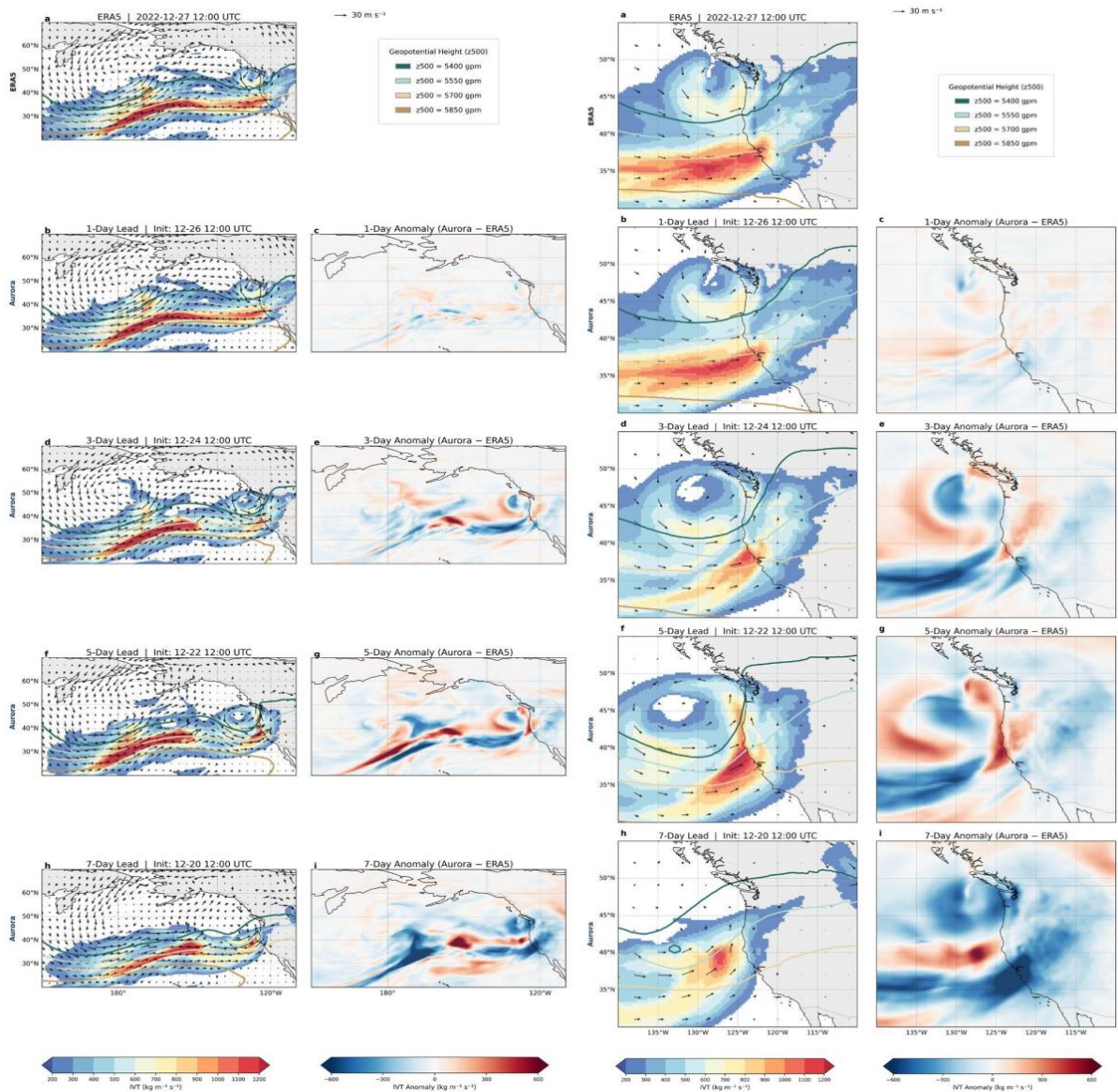

**Figure 5.** Pacific outlook and synoptic domain prediction of IVT for the December 2022 California atmospheric river, comparing ERA5 and Aurora forecasts at 1-, 3-, 5-, and 7-day lead times.

3.4.4 Summary and Operational Implications

Across the Pacific outlook domain, Aurora maintains high spatial coherence through 3–5 day lead times, but skill over the California region degrades more rapidly, where amplitude underestimation and positional errors reduce local IVT structure. This scale-dependent degradation reflects the nonlinear sensitivity of landfall impacts to modest upstream phase errors. The deterioration by 7 days over California implies a substantially shorter predictability horizon for local hazard intensity than for large-scale plume identification, aligning with broader forecast challenges observed for the December 2022/23 AR sequence. For example, experimental subseasonal forecasts skillfully captured the regime shift from dry to wet conditions in late December 2022 at 2–3 week lead times, but seasonal forecasts notably underpredicted precipitation magnitude, illustrating common limits of long-lead extreme precipitation prediction (DeFlorio et al., 2023).

From an operational perspective, these results suggest that reliable detection of large-scale AR presence may extend to medium-range lead times (3–5 days) for Aurora, consistent with dynamical numerical and subseasonal systems. However, quantitative assessment of landfall intensity and localized hazard potential becomes increasingly uncertain beyond these leads, mirroring findings in dynamical AR forecast research where subseasonal products capture regime shifts but often underpredict magnitude and timing of extreme precipitation (Schubert et al., 2024; DeFlorio et al., 2023).

Critically, the systematic underestimation of IVT magnitude at longer lead times underscores a consistent amplitude damping in Aurora forecasts, implying potential model or initialization biases in moisture transport strength. This characteristic bias structure may present opportunities for targeted correction strategies, such as bias adjustment, improved initialization, or controlled perturbation techniques, to better stabilize forecast amplitude and improve hazard estimates.

Overall, the AR case study demonstrates that while foundation model forecasts retain coherent dynamical structure at extended lead times, local hazard-relevant diagnostics (e.g., IVT intensity and precise plume orientation) deteriorate more rapidly. This underscores the importance of scale-aware verification in operational atmospheric river prediction: synoptic-scale presence signals can be informative early, but robust, quantitatively actionable hazard predictions at the regional level tend to be confined to shorter lead times that align with traditional synoptic and medium-range forecast skill envelopes.

**3.5 Extreme Precipitation**

Extreme precipitation predictability was assessed using two in-sample events: the July 2010 Pakistan floods and the August 2020 Sudan floods, and two out-of-sample events: the July 2021 Western Europe floods and the July 2022 Appalachian Floods.

3.5.1 Lead-Time Dependent Skill

**Table 7.** Extreme Precipitation Events Performance Summary

| Event | Lead Time | Prececiptation RMSE (mm) | Prececiptation Bias (mm) | Pattern Corr | Spatial Extent | IoU (1mm/6h threshold) |
|---|---|---|---|---|---|---|
| Pakistan 2010* | 1-day | 1.7967 | -0.1489 | 0.5552 | 0.1301 | 0.6927 |
|  | 3-day | 1.9621 | -0.1423 | 0.4481 | 0.1346 | 0.6356 |

|  | 5-day | 2.0394 | -0.1539 | 0.397 | 0.1307 | 0.5525 |
|---|---|---|---|---|---|---|
|  | 7-day | 2.2337 | -0.1475 | 0.2624 | 0.1279 | 0.5324 |
| Sudan 2020* | 1-day | 1.8606 | -0.1406 | 0.5357 | 0.12 | 0.2483 |
|  | 3-day | 1.9323 | -0.1617 | 0.4726 | 0.1185 | 0.2579 |
|  | 5-day | 2.0111 | -0.1718 | 0.4017 | 0.1156 | 0.2384 |
|  | 7-day | 2.1711 | -0.1608 | 0.2821 | 0.1231 | 0.1433 |
| Western Europe 2021 | 1-day | 3.1253 | -0.1335 | 0.3919 | 0.1211 | 0.3217 |
|  | 3-day | 3.2495 | -0.1382 | 0.283 | 0.1216 | 0.2237 |
|  | 5-day | 3.3166 | -0.1506 | 0.2187 | 0.1172 | 0.2785 |
|  | 7-day | 3.3835 | -0.1364 | 0.1601 | 0.1224 | 0.1209 |
| Appalachian 2022 | 1-day | 53.6356 | -0.1916 | 0.027 | 0.1182 | 0.2193 |
|  | 3-day | 53.6446 | -0.1817 | 0.0183 | 0.1269 | 0.1525 |
|  | 5-day | 53.6493 | -0.1692 | 0.0137 | 0.1342 | 0.1580 |
|  | 7-day | 53.6535 | -0.196 | 0.0094 | 0.1179 | 0.0489 |

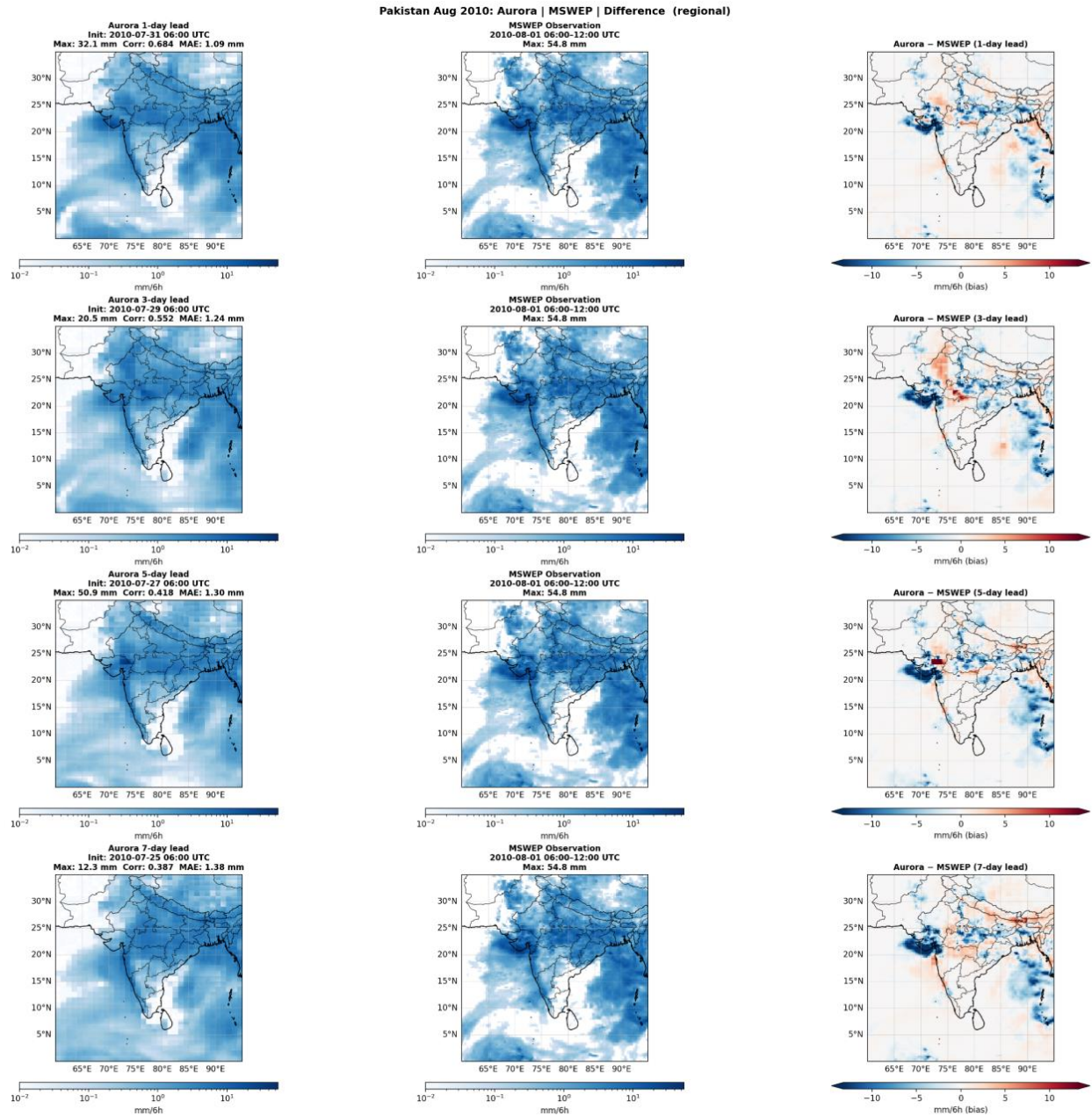

**Figure 6.** Precipitation prediction for extreme precipitation events, comparing MSWEP and Aurora forecasts at 1-, 3-, 5-, and 7-day lead times.

Across all four events, forecast skill degraded progressively with lead time, though the rate and character of degradation varied substantially between in-sample and out-of-sample cases. The two in-sample monsoon events (Pakistan 2010, Sudan 2020) showed the strongest short-range performance, with IoU above the 1 mm/6h threshold of 0.48 and 0.46 at 1-day lead, respectively, declining to 0.25 by 7-day lead for both events. Pattern correlations followed a similar trajectory, falling from 0.56-0.54 at 1-day lead to 0.26-0.28 at 7-day lead, indicating that Aurora gradually loses the ability to reproduce the spatial distribution of heavy rainfall as initialization distance increases. The two out-of-sample events (Western Europe 2021, Appalachian 2022) exhibited weaker initial skill and more pronounced degradation, with IoU declining from 0.30-0.31 at 1-day lead to 0.16-0.18 at 7-day lead, consistent with the greater predictive challenge posed by convective-dominated events in extratropical settings.

A clear in-sample versus out-of-sample contrast is evident in the short-range skill. For Pakistan 2010 and Sudan 2020, which fall within Aurora's ERA5 training period (1979-2020), pattern correlation at 1-day lead exceeded 0.53 and RMSE remained below 1.9 mm/6h. These values are notably higher than those for the two post-2020 events, where 1-day pattern correlation reached only 0.39 for Western Europe 2021 and 0.03 for Appalachian 2022. While the Appalachian result is likely partly attributable to the mesoscale convective nature of that event, the broader performance gap between in-sample and out-of-sample cases suggests some degree of overfitting to the ERA5 climatological distribution of precipitation patterns.

Aurora showed its strongest precipitation forecast skill at short lead times, with performance generally highest at 1-day lead and then degrading progressively through 7 days. Among the four cases, the best short-range performance was obtained for the Pakistan Aug 2010 and Sudan Aug 2020 events. For Pakistan, the 1-day forecast yielded an RMSE of 1.7967 mm, a bias of -0.1489 mm, a pattern correlation of 0.5552, and an IoU of 0.4801. Sudan showed similarly strong performance, with a 1-day RMSE of 1.8606 mm, bias of -0.1406 mm, pattern correlation of 0.5357, and IoU of 0.4647. In both cases, the forecast and observed spatial extents were also close: 0.1301 versus 0.1345 for Pakistan, and 0.1200 versus 0.1280 for Sudan. These results indicate that Aurora was able to capture both the broad spatial distribution and the areal coverage of extreme precipitation reasonably well at short leads for these two events.

Performance for the Western Europe Jul 2021 event was weaker, though still retained some useful spatial skill at 1-day lead. RMSE was higher at 3.1253 mm, while pattern correlation and IoU were 0.3919 and 0.3025, respectively. Interestingly, the forecast extent of 0.1211 exceeded the observed extent of 0.0858, indicating that Aurora overestimated the areal coverage of the event even while maintaining a slightly negative mean bias (-0.1335 mm). This suggests that the model spread precipitation too broadly in space while underestimating local intensity on average.

The Appalachian Jul 2022 event was by far the most difficult case. Although IoU at 1-day lead was 0.3091, the pattern correlation was only 0.0270, indicating almost no meaningful reproduction of the observed spatial structure. In addition, RMSE was extremely large at 53.6356 mm, far exceeding the values for all other events. This combination of very high RMSE and near-zero pattern correlation suggests that Aurora struggled substantially with both precipitation magnitude and spatial placement in this case. The forecast extent (0.1182) was also notably higher than the observed extent (0.0869), implying that the model again produced a precipitation area that was too broad.

3.5.2 Spatial Structure and Systematic Biases

Across all events, Aurora exhibited a consistent dry bias in regional mean precipitation (-0.13 to -0.19 mm/6h), indicating systematic underestimation of precipitation intensity. Paradoxically, this dry bias coexisted with spatial overestimation: for the Western Europe 2021 and Appalachian 2022 events, forecast spatial extent at the 1 mm/6h threshold exceeded the observed extent (0.12 versus 0.09 for both events), while RMSE remained elevated. This combination — spatially too broad but locally too weak — is a characteristic fingerprint of AI weather models applied to convective-scale events: the synoptic-scale moisture convergence that favors precipitation is captured, but the mesoscale convective organization responsible for localized extreme accumulations is not resolved.

Pattern correlation retained moderate predictive value through 5-day lead for the Pakistan and Sudan events (0.40 at 5-day lead for both), indicating that Aurora reproduces the large-scale circulation patterns governing monsoon precipitation with reasonable fidelity across the medium range. This is qualitatively similar to the behavior observed for freeze and heatwave events in sections 3.2-3.3, where pattern correlation remained elevated even as threshold exceedance skill degraded. In contrast, the near-zero pattern correlation for Appalachian 2022 across all lead times (0.027 at 1-day lead, declining to 0.009 at 7-day lead) indicates that this event was driven by mesoscale processes entirely below Aurora's effective predictability scale, and the moderate IoU values (0.31-0.18) reflect overlap driven by the broad climatological wetness of the region rather than genuine dynamical skill.

### 3.5.3 Operational Implications

Aurora provides operationally useful short-range guidance for large-scale monsoon precipitation events at 1–3 day leads, with IoU above 0.38 and pattern correlation above 0.45 for Pakistan 2010 and Sudan 2020. For these synoptically driven events, the model captures the regional distribution of heavy rainfall well enough to support broad-area flood watches, though the decoder tends to smooth the upper tail—producing less "peaky" rainfall fields and under-representing localized maxima—so impact-relevant intensity thresholds should be recalibrated when interpreting totals. For convective flash flood events such as Western Europe 2021 and Appalachian 2022, short-range spatial skill is marginal to poor, and Aurora output alone is insufficient for localized early warning. Post-processing, statistical downscaling, or coupling with a convection-permitting model would be required to translate Aurora's synoptic-scale signal into actionable precipitation guidance at watershed scales.

An additional caveat is that precipitation is not a native prognostic variable in Aurora, and is instead produced through a decoder, introducing extra uncertainty beyond the core dynamical forecast. This suggests a clear path for improvement: a precipitation-aware, fine-tuned Aurora variant could reduce this limitation by learning event-dependent rainfall structures directly within the model's training procedure—particularly in the extremes—whereas the Lehmann-style decoder does not explicitly emphasize or preserve extreme-rainfall behavior.

Overall, precipitation remains a more challenging predictability target than temperature extremes or tropical cyclone tracks, and the results presented here suggest that Aurora's practical utility for flood forecasting is currently concentrated at the synoptic and regional spatial scales consistent with its training, rather than at the localized scales required for flash-flood guidance.

**4 Conclusions**

This study presents a cross-event evaluation of Aurora's predictability across multiple classes of extreme weather using an event-based diagnostic framework. Across tropical cyclones, temperature extremes, atmospheric rivers, and extreme precipitation events, Aurora exhibits clear strengths at short-to-medium lead times and systematic limitations at longer leads.

For tropical cyclones, Aurora provides operationally useful track guidance at 1–3 day leads, with mean position errors of 20–60 km for most systems and low initialization sensitivity. However, intensity biases and catastrophic failures in recurving cases (e.g., Hinnamnor) highlight limitations in representing midlatitude interaction and storm structure. Aurora's short-range TC skill is competitive with other AI models, but does not yet match the consistency of operational dynamical–statistical consensus systems across all scenarios.

For temperature extremes, Aurora shows strong skill at 1–7 day leads for both freeze and heatwave events, with high pattern correlation and spatial overlap. Beyond ~7–10 days, however, threshold-based skill collapses while large-scale circulation patterns remain moderately skillful. This divergence, accurate synoptic structure but regressed surface amplitude, indicates that Aurora captures blocking and large-scale flow regimes but fails to sustain thermodynamic intensity at subseasonal leads.

For atmospheric rivers, short-range (1–3 day) IVT structure and landfall location are well reproduced. Skill degradation at 5–7 days occurs primarily through systematic underestimation of moisture transport intensity and growing regional displacement errors, especially at landfall. Local hazard-relevant diagnostics degrade more rapidly than large-scale plume identification.

Across event types, a common subseasonal failure mode emerges: moderate skill in large-scale dynamical fields (e.g., Z500, MSLP) but collapse of threshold exceedance skill at 14–21 day leads. This behavior is consistent with intrinsic predictability limits of midlatitude atmospheric dynamics rather than solely model-specific deficiencies. From an operational perspective, Aurora's rapid inference and competitive short-range skill position it as a complementary tool for:

- Rapid guidance and ensemble design
- Short-range decision support (1–7 days)
- Synoptic regime identification at longer leads

However, bias correction, ensemble generation, and hybrid ML–physics integration are necessary before standalone operational deployment for impact-level forecasting. Aurora's limitations—particularly amplitude damping and event-specific variability—underscore the importance of combining data-driven models with physically based approaches.

Ultimately, foundation models trained on reanalysis inherit the fundamental predictability limits of the atmosphere. Aurora's greatest value lies in fast, skillful short-to-medium range guidance, while reliable deterministic extreme-event forecasting beyond ~7–10 days remains constrained by atmospheric dynamics.

**Conflict of Interest Disclosure**

The authors declare that they have no conflict of interest.

**Data Availability Statement**

ERA5 reanalysis data used for model initialization and verification were obtained from the Copernicus Climate Data Store (CDS), hosted by the European Centre for Medium-Range Weather Forecasts (ECMWF): https://cds.climate.copernicus.eu/

The Aurora pretrained model used in this study is publicly available through Microsoft's official GitHub repository: https://github.com/microsoft/aurora

Precipitation fields were generated using the auxiliary Aurora precipitation decoder described by Lehmann et al. (2025), which maps Aurora's latent atmospheric state representation to precipitation without modifying the pretrained model's dynamical core. The decoder is available at: https://github.com/lehmannfa/aurora-lite-decoder

MSWEP Version 3 (Multi-Source Weighted-Ensemble Precipitation; hourly 0.1° resolution, 1979–present) is available via its dataset documentation and associated preprint: Wang et al. (2026), https://doi.org/10.48550/arXiv.2602.01436

IBTrACS tropical cyclone best-track data were obtained from the International Best Track Archive for Climate Stewardship (IBTrACS), maintained by NOAA National Centers for Environmental Information: https://www.ncei.noaa.gov/products/international-best-track-archive

All processed data, derived metrics, and event-based evaluation outputs generated in this study are available from the corresponding author upon reasonable request.

Doss-Gollin, J., Farnham, D. J., Lall, U., & Modi, V. (2021). How unprecedented was the February 2021 Texas cold snap? *Environmental Research Letters*, *16*(6), 064056. https://doi.org/10.1088/1748-9326/ac0278

European Centre for Medium-Range Weather Forecasts (ECMWF). (2019). *ERA5 Reanalysis (ECMWF Reanalysis v5).* Copernicus Climate Change Service (C3S) Climate Data Store. Accessed May 27, 2025. https://www.ecmwf.int/en/forecasts/dataset/ecmwf-reanalysis-v5

Food and Agriculture Organization (FAO). (2020). *The Sudan 2020 Flood Impact Rapid Assessment, September 2020 – Rapid Assessment Report.* FAO, Rome. https://www.fao.org/fileadmin/user_upload/emergencies/docs/South%20Sudan_Flood%20Assesment%20Report.pdf

Gahtan, J., Sivakumar, A., Gerritsma, M. I., Oosterlee, C. W., Peixoto, P. S., & Sanderson, R. (2024). Machine learning for tropical cyclone track forecasting: Investigations with Tempest Extremes and the MPAS model. *Artificial Intelligence for the Earth Systems*, *3*(4), e230099. https://doi.org/10.25921/82ty-9e16

Gupta, A., Sheshadri, A., & Suri, D. (2025). *MAUSAM: An observations-focused assessment of global AI weather prediction models during the South Asian monsoon*. arXiv. https://doi.org/10.48550/arXiv.2509.01879

Houze, R. A., Rasmussen, K. L., Medina, S., Brodzik, S. R., & Romatschke, U. (2011). Anomalous Atmospheric Events Leading to the Summer 2010 Floods in Pakistan. *Bulletin of the American Meteorological Society*, *92*(3), 291–298. http://www.jstor.org/stable/26226849

Huang, J., Hitchcock, P., Tian, W., &Sillin, J. (2022). Stratospheric influence on the development of the 2018 latewinter European cold air outbreak.Journal of Geophysical Research:Atmospheres, 127, e2021JD035877. https://doi.org/10.1029/2021JD035877

India Meteorological Department (IMD). (2020). *Super Cyclonic Storm "AMPHAN" over Southeast Bay of Bengal (16–21 May 2020) — Summary.* Government of India / IMD (published on ReliefWeb). URL: https://reliefweb.int/report/india/super-cyclonic-storm-amphan-over-southeast-bay-bengal-16th-21st-may-2020-summary

Knapp, K. R., Kruk, M. C., Levinson, D. H., Diamond, H. J., & Neumann, C. J. (2010). The International Best Track Archive for Climate Stewardship (IBTrACS): Unifying tropical cyclone data. *Bulletin of the American Meteorological Society*, *91*(3), 363–376. https://doi.org/10.1175/2009BAMS2755.1

Lam, R., Sanchez-Gonzalez, A., Willson, M., Wirnsberger, P., Fortunato, M., Alet, F., … & Battaglia, P. W. (2023). Learning skillful medium-range global weather forecasting. *Science*, *382*(6677), 1416–1421. https://doi.org/10.1126/science.adi2336

Lang, S., Alexe, M., Chrust, M., Dramsch, J., Pinault, F., Raoult, B., … & Dueben, P. (2025). AIFS - ECMWF's data-driven forecasting system. *ECMWF Technical Memorandum*, 920.
34